\documentclass[12pt]{article}
\usepackage{graphicx}
\begin{document}
\title{A weak instability in an expanding universe?}
\author{\normalsize Bob
Holdom\thanks{bob.holdom@utoronto.ca} \\
\small {\em Department of Physics, University of Toronto}\\\small {\em
Toronto, Ontario,}
M5S1A7, CANADA}\date{}\maketitle
\begin{picture}(0,0)(0,0)
\put(310,205){UTPT-02-02}
\put(310,190){hep-th/0203087}
\end{picture}
\begin{abstract}  
We use higher derivative classical gravity to study the nonlinear coupling between the cosmological
expansion of the universe and metric oscillations of Planck frequency and very small
amplitude. We derive field equations at high orders in the derivative expansion and find that the
nature of the new dynamics is extremely restricted. For the equation of state parameter $w>0$ the
relative importance of the oscillations grows logarithmically. Their effect on the cosmological expansion 
resembles that of dark energy.
\end{abstract}
\baselineskip 18pt
\section{Introduction}
A general gravitational action can be arranged as a derivative expansion, where higher derivative
terms contain more factors of the curvature tensor and its covariant derivatives. As such
this could be viewed
as the expansion of some underlying theory. But major problems arise if
some truncation of the derivative expansion is treated as a fundamental theory.
For example the inverse propagator of the high derivative theory is at best of the form
\begin{equation}k^2\prod\limits_{i}^{} ( {k}^{2}-{m}_{i}^{2} )\;\;\;{\rm with}\;{\rm all}\;{m}_{i}^{2}  >  0
.\end{equation}
Close to every other zero the $k^2$ will have the wrong sign. These ghost poles lead to
loss of unitarity, the possibility of unstable solutions, etc.
At a deeper level the $m_i$ probably have little to do with the 
spectrum of the underlying theory. Also a truncated derivative expansion 
will have solutions bearing 
no resemblance to the solutions of the true theory.
Higher derivative gravity only makes sense as an effective field theory where the 
higher derivative terms are properly treated as perturbations \cite{e}.
(For a contrary view see \cite{d}.)

Nevertheless we will show in this paper that higher derivative 
gravity in the case ${m}_{i}^{2}  >  0$ can
serve as an instructive and suggestive framework for the investigation
of the following question.
Can there be a dynamical nonlinear coupling between physics at very large and very small time scales 
(UV-IR coupling) in a theory of gravity?

We will consider a class of solutions to higher derivative gravity theories of the FRW form, 
where the standard FRW
time dependence for the scale factor can be recovered as a special case. This special case corresponds to
particular initial conditions, and departures from this will
yield oscillations with Planck scale frequencies in the scale factor of the FRW metric. 
It has been noted before \cite{f} that
rapid  metric oscillations can emerge from higher derivative theories.
We will find that the oscillations, even when averaged over, produce a 
departure from the FRW solution, where the
fractional amount of this departure grows with time. We refer to 
this as a \textit{weak} dynamical instability
since the growth of these oscillations turns out to be logarithmic.

We claim that this is a generic property of higher derivative gravity, true at any order in the 
derivative expansion.
We have reached this conclusion through a study of the field equations at high order in the derivative
expansion. 
Since Planck frequencies are involved, derivatives of the oscillating component are not suppressed.
The amplitude of the oscillation on the other hand is extremely small, at least at late times, if
the impact of the oscillations on the equations is smaller than or of order the usual effects
of cosmological expansion. We therefore consider a small amplitude expansion, where terms quadratic
in the scale factor and its derivatives produce the leading nonlinear effects.
It is these nonlinear effects that couple the oscillations and cosmological expansion together 
and give rise to the dynamical instability.

As we have said the new solutions can be arbitrarily close to the standard FRW solution, in which
case there will be an arbitrarily long time where the effects of all higher derivative terms are 
much smaller than the Einstein
curvature term. But eventually the effects of higher powers of curvature can become as important
as the Einstein term, and the effect on the cosmological expansion can be significant.

One would naively think that the large and growing number of parameters in higher derivative actions
would produce results that are very ambiguous. On the contrary we find that
in the small amplitude expansion that the derivative expansion is extremely restrictive.
Each order in the derivative expansion, at six derivatives and beyond, results in 
only two parameters, one of which multiplies
both the leading and next-to-leading terms in the small amplitude expansion. And then quite
surprisingly we find that the second parameter drops out of the quantities of interest.
Thus a very particular pattern emerges in the derivative expansion that allows us to relate
the amplitudes and energies of the oscillating modes to the matter sources.

We start with the gravitational action
\begin{equation}
S  =  {M}_{\rm{Pl}}^{2}\int_{}^{}{d}^{4}x\sqrt {-g}( -2\Lambda +R
+a{R}^{2}+b{R}_{\mu \nu }{R}^{\mu \nu }+c{R}_{\mu \nu \lambda \kappa }{R}^{\mu \nu \lambda \kappa }+... )\end{equation}
where $a$, $b$, and $c$ are of order $M_{\rm Pl}^{-2}$. In this work we will only consider a spatially
flat metric,
\begin{equation}d{s}^{2}  =  -d{t}^{2}+a(t)^2{\delta }_{ij }d{x}^{i}d{x}^{j}.\label{e1}\end{equation}

The field equations are
\begin{equation}\hat{G}_{\mu \nu }={T}_{\mu \nu }+\Lambda {g}_{\mu \nu }\label{e7} \end{equation}
where $8\pi G$ has been absorbed into ${T}_{\mu \nu }$ and $\Lambda$.
$\hat{G}_{\mu \nu }$ is a power series in derivatives, where the two derivative terms
 make up the Einstein tensor.
The matter described by ${T}_{\mu \nu }$ has the equation of state $p(t)=w\rho(t)$. 
We introduce a dimensionless Planckian time
\begin{equation}\tilde{t}  =  {\frac{t}{\sqrt {6a+2b+2c}}}  \approx   {M}_{{\rm Pl}}\:t,\;\;\;A( \tilde{t} )  \equiv   2\ln(a( t) )
\label{e6},\end{equation}
and in the following we drop the tilde. For the above metric, $6a+2b+2c$ is the only 
combination of the four-derivative parameters
that appear in the field equations, and we must assume that it is positive.
We absorb a further factor of $6a+2b+2c$ into the terms in (\ref{e7}), thus making
${T}_{\mu \nu }$ and $\Lambda$ dimensionless.

In terms of this dimensionless time
we will be considering oscillatory behavior for $A(t)$ with roughly unit frequencies and very small amplitude
$\varepsilon$. Thus ${\partial}^{n}A( t )/\partial{t}^{n}  \approx   \varepsilon$ and
we consider a small amplitude expansion in powers of $\varepsilon $. The linear terms are contained in $\hat{G}_{ii}$.
\begin{equation}\hat{G}_{00}[A( t )]  =  {\cal O}( {\varepsilon }^{2} )\end{equation}
\begin{equation}\hat{G}_{ii}[A( t )]  =  -\sum\limits_{n\ge 1}^{} {c}_{2n}{\partial }_{t}^{2n}A( t )
+{\cal O}( {\varepsilon }^{2} )\end{equation}
The definitions above yield $c_2=c_4=1$. At 6 derivatives and above these linear terms
originate from terms in the action containing two curvature tensors, or contractions thereof, and the appropriate number of
covariant derivatives.

To describe our results at higher orders in $\varepsilon$ we make the following definitions,
\begin{equation}\hat{G}_{00}[A( t )]  \equiv  \sum\limits_{n}^{} {{c}_{2n}{\cal G}}_{\rho }^{( 2n )}
\label{e8}\end{equation}
\begin{equation}\hat{G}_{ii}[A( t )]  \equiv  -\sum\limits_{n}^{} {{c}_{2n}{\cal G}}_{p }^{( 2n )}
\label{e9}\end{equation}
where ${\cal G}_{\rho }^{( 2n )}$ and ${\cal G}_{p}^{( 2n )}$ contain $2n$ derivatives.

We display in the Appendix the results
up to 12 derivatives, for ${{\cal G}}_{p }^{(2n)}$ up
to ${\cal O}( {\varepsilon }^{2} )$ and ${{\cal G}}_{\rho }^{(2n)}$ up
to ${\cal O}( {\varepsilon }^{3} )$. 
As the number of derivatives grows the number of possible terms in the action
that contribute grows very fast.
Despite this we see that in addition to $c_{2n}$ only one new parameter ${\chi }_{2n}$ appears at each
order in the total number of derivatives, to the order in $\varepsilon$ displayed.
(Actually at 2 and 4 derivatives there are no parameters given the definitions above and the terms
at all orders in $\varepsilon$ are displayed.)

These results were obtained without obtaining the general field equations for general metric.
Since the $t$ dependence of the metric (\ref{e1}) appears as a general function of $t$ in ${g}_{ii}$,
we can insert the metric (\ref{e1}) into the action and vary with respect to $A(t)$. This will give the
$ii$ field equation, and thus the ${{\cal G}}_{p}^{( 2n )}$,
 for this particular metric. The ${{\cal G}}_{\rho}^{( 2n )}$ can then be obtained 
through the use of the conservation equation (generalized Bianchi identity)
${D}^{\mu }\hat{G}_{\mu \nu }  =  0$, which we satisfy to ${\cal O}( {\varepsilon }^{3} )$. 
In fact this completely determines ${{\cal G}}_{\rho}^{( 2n )}$
at order ${\cal O}( {\varepsilon }^{m+1} )$ given ${{\cal G}}_{p}^{( 2n )}$
at order ${\cal O}( {\varepsilon }^{m} )$. All the terms in the field equations, to any order in $\varepsilon $
and for any given
action, can be obtained in this way, but only the terms at the orders in $\varepsilon $ relevant for our study are given
in the Appendix.

\section{Pure oscillations}
We start with the oscillating vacuum solutions of the linearized equations, corresponding to
the Lagrangian
 \begin{equation}{\cal L}  =  -{\frac{1}{2}}A( t )(\sum\limits_{n\ge 1}^{} c_{2n} {\partial }_t^{2n}) A( t )
.\end{equation}
For $A(t)$ of the form
 ${\rm cos}( \omega t )$ the allowed frequencies must satisfy
\begin{equation}\sum\limits_{n\ge1}^{} {(-1 {)}^{n }c}_{2n}^{ }\:{\omega }^{2n}  =  0
\label{e5}.\end{equation}
When a negative or complex ${\omega }^{2}$ solves (\ref{e5}) the theory is exponentially unstable, a case
we will not consider further. For all roots, besides the vanishing one, to be real and positive
  it is necessary (but not sufficient) that all ${c}_{2n}>0$.

Given the set of allowed frequencies $\omega_i$ we can write the Lagrangian as
\begin{equation}{\cal L}  =  -{\frac{1}{2}}A( t )\left[\prod\limits_{i\ge 0}^{} ( {\partial }_t^{2}+{\omega }_{i}^{2})\right] A( t )
,\end{equation}
after absorbing a constant normalization into $A(t)$.
We will assume that $\omega_j>\omega_i$ for $j>i$, and $\omega_0=0$.
This can be rewritten as a sum of conventional free field terms by defining\footnote{
This generalizes the definitions in \cite{d}.}
\begin{equation}{\phi }_{i}( t )  =  {\frac{\left[{\prod\limits_{j\ne i}^{} ( {\partial }_t^{2}+{\omega }_{j}^{2} )}\right]A(  t  )}{{\left|{\prod\limits_{j\ne i}^{} ( {\omega }_{i}^{2}-{\omega }_{j}^{2} )}\right|}^{1/2}}}
.\end{equation}
Note that $A(t)\sim \cos(\omega_i t)$
corresponds to all $\phi_j(t)$ vanishing except for $\phi_i(t)$. The Lagrangian then becomes
\begin{equation}{\cal L} =  -{\frac{1}{2}}\sum\limits_{i\ge 0}^{} (- {)}^{i}{\phi }_{i}( t )( {\partial }_t^{2}+{\omega }_{i}^{2} ){\phi }_{i}( t )
.\end{equation}
The $(-)^i$ factor means that every other term comes in with the wrong sign, and in particular
this is the case  for 
the mode with the lowest nonvanishing frequency $\omega_1$. The wrong sign in the Lagrangian
means that the classical energy of these modes is negative.  The energy density of the $i$th oscillating
mode with amplitude $\varepsilon_i$ is thus proportional to $(-)^i \varepsilon_i^2$.
We will refer to this as a gravitational energy. These minus signs appearing in the linearized
theory cause problems in the attempt
to quantize higher derivative gravity.

These vacuum oscillating solutions of the linearized theory are of course not solutions of the full
nonlinear theory. In particular an oscillating $A(t)$ will generate non-oscillating terms in the 
field equations at ${\cal O}( {\varepsilon }^{2} )$,
and a source ${T}_{\mu \nu }$ must contribute at this order for a solution to exist.
We will include both a cosmological constant
$\Lambda$ and matter with equation of state $p(t)=w \rho(t)$.
 With $A(t)$ in the form
\begin{equation}A( t )  =  \varepsilon  {\rm cos}( \omega \:t)+{a}_{2}{ \varepsilon }^{2}{\rm cos}( 2\:\omega \:t )+...
\end{equation}
we expand the field equations to ${\cal O}( {\varepsilon }^{2} )$
 using the results for ${\cal G}_{\rho }^{( 2n )}$ and ${\cal G}_{p}^{( 2n )}$ in the Appendix.
 A solution would consist in relating
$\varepsilon $ and $a_2$ to the gravitational sources.
Solutions at higher order in $\varepsilon $ could be found by continuing the expansion
of $A(t)$ in the obvious way.\footnote{Solutions at ${\cal O}( {\varepsilon }^{3} )$ were found
in the $2+4$ derivative case in \cite{c}.}

Despite the complexity of the expressions in the Appendix, we find very simple results
for the oscillating $A(t)$,
and a clear pattern emerges in the terms with 2 through to 12 derivatives. We assume
that this pattern will continue at yet higher derivatives.
\begin{equation}{\varepsilon }^{2} {3\over8}({\cal B}+{\cal A}{\rm c}{\rm o}{\rm s}( 2\omega t ) )  =  \rho ( t )+\Lambda 
\label{e2}\end{equation}
\begin{equation}{-\varepsilon {\cal A}{\rm c}{\rm o}{\rm s}( \omega t )+\varepsilon }^{2}( {3\over8}{\cal A}+{\cal C}{\rm c}{\rm o}{\rm s}( 2\omega t ) )  =  p ( t )-\Lambda \end{equation}
\begin{equation}{\cal A}=\sum\limits_{n\ge 1}^{} ( - {)}^{n}{c}_{2n}{\omega }^{2n}\end{equation}
\begin{equation}{\cal B}=\sum\limits_{n\ge 1}^{} ( - {)}^{n-1}( 2n-1 ){c}_{2n}{\omega }^{2n}\end{equation}
\begin{equation}{\cal C}={\cal C}(a_2,c_{2n},\chi_{2n},\omega)\end{equation}
We see that $\rho(t)$ at ${\cal O}( {\varepsilon }^{2} )$ is completely determined 
by the same constants $c_i$ that appear at linear order; the additional
$\chi_{2n}$ parameters have all dropped out
of this equation.
The terms linear in $\varepsilon$ in the $p(t)$ equation must vanish separately,
thus determining the allowed frequencies as before:
 $\omega  \in  \{{\omega }_{i}\}$ such that ${\cal A}=0$. Thus at ${\cal O}( {\varepsilon }^{2} )$ 
we find that $\rho(t)$ is a constant.

In the $p(t)$ equation only the
${\cal C}\cos(2\omega t)$ term remains, which depends linearly on $a_2$ and the $\chi_{2n}$ parameters.
The solution, to be compatible with $p=w\rho$, is thus completed by 
choosing $a_2$ so that ${\cal C}=0$. In this way the $\chi_{2n}$ parameters
affect $A(t)$ only at order ${\cal O}( {\varepsilon }^{2} )$, through the value of $a_2$. The result is that
the total pressure, $p(t)-\Lambda$, vanishes and thus \begin{equation}\rho(t)=\Lambda/w.\end{equation}
Then to obtain a solution for the mode with frequency $\omega_i$, its amplitude $\varepsilon_i$ 
is determined from (\ref{e2}) as
\begin{equation}{\varepsilon }_i^{2}{\frac{3}{8}}\sum\limits_{n\ge 1}^{} ( - {)}^{n-1}( 2n-1 ){c}_{2n}{\omega }_i^{2n}  =  \rho(1+w)
.\label{e4}\end{equation}

We need the value of ${\cal B}=\sum ( - {)}^{n-1}( 2n-1 ){c}_{2n}{\omega }^{2n}$ for the
various roots of ${\cal A}=\sum ( - {)}^{n}{c}_{2n}{\omega }^{2n}$. It turns out that ${\cal B}$ also
has a set of positive roots that are interlaced with those of ${\cal A}$. The result is that the sign of
${\cal B}$ alternates at each successive root of
${\cal A}$, with the implication  (assuming $w>-1$) that
\begin{equation}{\rm sign}(\rho)=(-)^i
\end{equation}
for the mode with frequency $\omega_i$. Thus the same modes that have negative gravitational energy also require a negative matter energy
for their existence. In other words for these modes
$T_{00}+\Lambda g_{00}<0$ whereas the modes with positive gravitational energy have $T_{00}+\Lambda g_{00}>0$.
In both cases $T_{ii}+\Lambda g_{ii}=0$. Thus in order not to violate positive energy conditions
for matter, we may conclude that the set of
oscillating solutions to the full theory does not include the modes with negative gravitational energy. 

Note that the vanishing of $p(t)-\Lambda$ with $\Lambda\ne 0$ requires $w\ne 0$.
 On the other hand if
$\Lambda=0$ then the same analysis would yield solutions for $w=0$. Thus
pressureless matter can act as a source of the oscillations, with the amplitude of the oscillations determined by
the energy density. In this case $\rho$ can be freely chosen, unlike the case with $\Lambda\ne 0$.

Energy conservation for matter implies that $\rho(t)$ must oscillate in an oscillating background,
since $\rho ( t )a( t {)}^{3( 1+w )}$ (where $a(t)=e^{A(t)/2}$ is the scale factor) must be a constant. 
Thus an oscillating component of $\rho(t)$
will occur at ${\cal O}(\varepsilon^3)$. In summary an oscillation amplitude $\varepsilon$ of $A(t)$ 
corresponds to a source of
${\cal O}(\varepsilon^2)$ having an oscillating component of ${\cal O}(\varepsilon^3)$,
a situation very different than the pure linearized theory. The
implications of any of this for the development of a quantum theory will not be pursued here.

\section{Oscillations and expansion}
We turn now to consider the dynamics of these nonlinear oscillations in the context of the expanding
universe. We will find that modes that were prohibited in the pure oscillating case become involved in a
weak instability in the expanding case. This dynamics depends on the equation of state.

We will have to explore this numerically in theories truncated to various finite numbers of derivatives.
We use the equation
of state to eliminate $\rho(t)$ and $p(t)$ and obtain an equation for $A(t)$ alone.
A single equation can be dealt with quite successfully numerically, even though it is
nonlinear with high derivatives.
Once $A(t)$ is obtained then the other equation determines $\rho(t)$.
A good check on the accuracy of the numerical integration is provided by
the quantity $\rho ( t )a( t {)}^{3( 1+w )}$, which should remain constant.
We have confirmed this for all the cases studied.

\begin{figure}
\begin{center}
\includegraphics{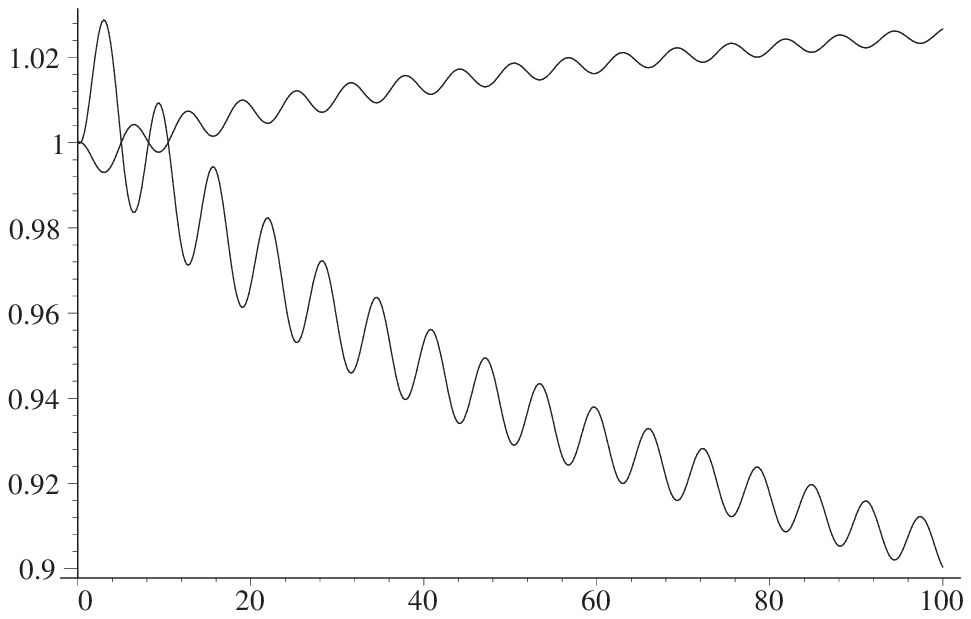}
\begin{picture}(0,0)(0,0)\large 
\put(0,165){$a(t)\over a(t)_{\rm FRW}$}
\put(0,20){$\rho(t)\over \rho(t)_{\rm FRW}$}
\end{picture}
\end{center}
\vspace{3ex}
\begin{center}
\includegraphics{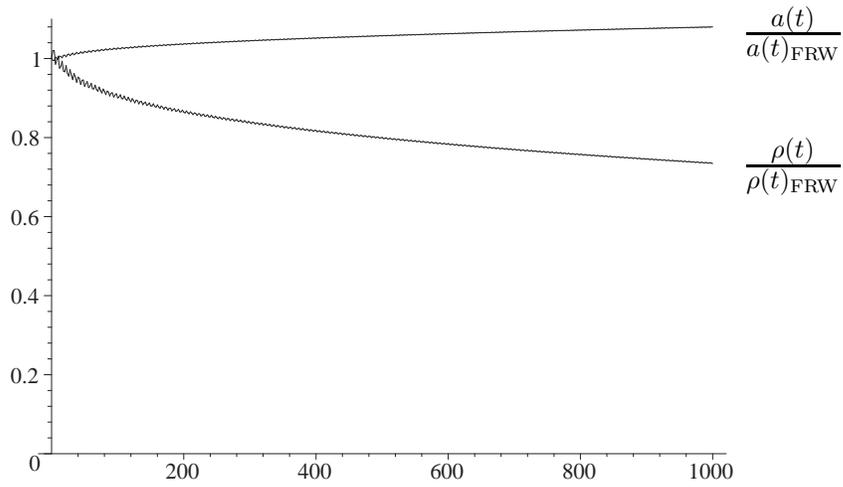}
\begin{picture}(0,0)(0,0)\large 
\put(0,165){$a(t)\over a(t)_{\rm FRW}$}
\put(0,115){$\rho(t)\over \rho(t)_{\rm FRW}$}
\end{picture}
\end{center}
\caption{These two plots display the same evolution for different ranges of $t$.
The $t$ here is actually $\tilde t-30$ for $\tilde t$ defined in (\ref{e6}).}
\label{f2} 
\end{figure}

We begin by truncating the equations at 4 derivatives and setting $\Lambda=0$.
We initially consider radiation domination $w=1/3$. Here it is easy
enough to keep all the terms, to all orders in $\varepsilon$.
A range of initial conditions gives rise to the positive energy expanding solutions of
interest to cosmology. In Fig.~(\ref{f2}) we plot 
$a(t)/a(t)_{\rm FRW}$ and $\rho(t)/\rho(t)_{\rm FRW}$ for different ranges of $t$.
For different initial conditions the size of the oscillating effect changes and the departure from the
FRW solution can in particular be made much smaller, but the qualitative form of the plots remains
the same. We will discuss the initial conditions further below; but for now we see that 
$a(t)$ increases faster, and $\rho(t)$ falls faster, than in the FRW case.

\begin{figure}
\begin{center}
\includegraphics{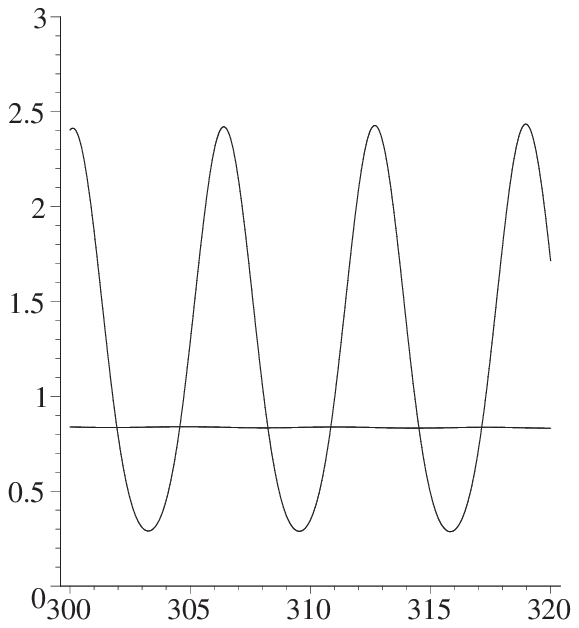}\hspace{3ex}
\includegraphics{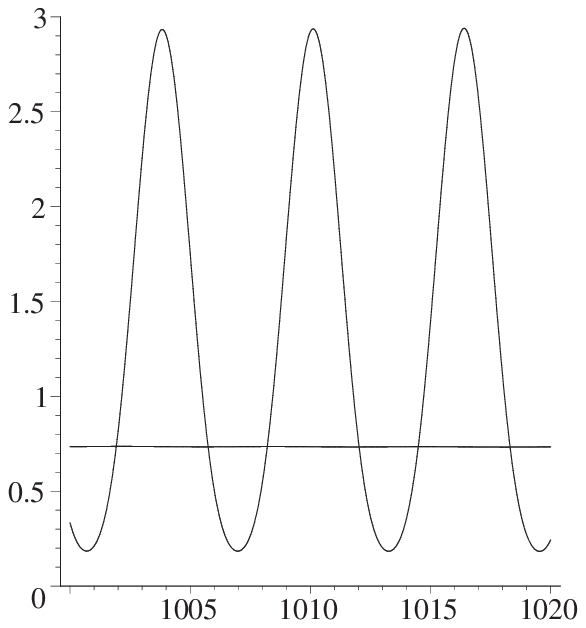}
\begin{picture}(0,0)(0,0)\large
\put(0,47){$\rho(t)\over\rho(t)_{\rm FRW}$}
\put(0,140){${H}^{2}(t)\over{H}(t)^{2}_{\rm FRW}$}
\put(0,50){\vector(-1,0){10}}
\put(0,145){\vector(-1,0){30}}
\end{picture}
\end{center}
\caption{These two figures show the relative growth of the oscillations.}
\label{f4} 
\end{figure}

We can consider the departure from the FRW equation $3{H}^{2}  =  \rho$ as an indication
of the importance of the higher derivative terms.
This is seen in the plot of ${H}^{2}(t)/{H}(t)^{2}_{\rm FRW}$ and $\rho(t)/\rho(t)_{\rm FRW}$
in Fig.~(\ref{f4}). We see the relative
smoothness of $\rho(t)$ and the growing oscillations in ${H}^{2}(t)/{H}(t)^{2}_{\rm FRW}$,
which implies the growing relative importance of the higher derivative terms. It is in this
sense that the oscillations are growing and signaling
a dynamical instability. We find that the dynamics is such that departure of 
$\rho(t)/\rho(t)_{\rm FRW}$ from unity grows like a power of $\ln(t)$,
and thus the instability is weak and physically interesting.

\begin{figure}
\begin{center}
\includegraphics{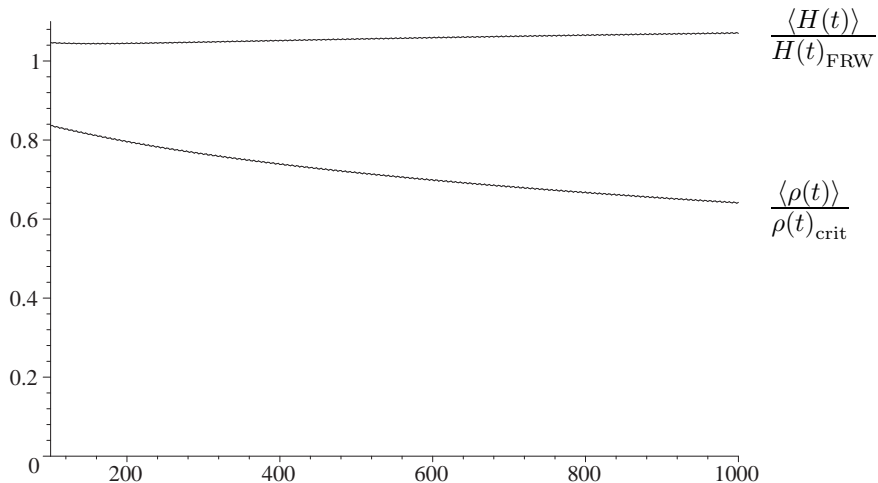}
\begin{picture}(0,0)(0,0)\large 
\put(0,100){$\langle \rho ( t ){\rangle }^{}\over\rho ( t {)}_{{\rm crit}}$}
\put(0,165){$\langle H(t)\rangle \over H ( t {)}_{{\rm FRW}}$}
\end{picture}
\end{center}
\caption{The bottom curve shows the smeared matter energy density $\langle \rho ( t ){\rangle }$
 compared with
the apparent critical value for a flat universe, where the latter is determined by 
$\langle H(t)\rangle$.}
\label{f5} 
\end{figure}

Since the oscillations will not be directly observed it makes sense to consider quantities that are
 suitably time averaged, so as to smear out the oscillations.
For example the apparent critical energy density of the universe will be given by 
${\rho ( t )}_{{\rm crit}}  \equiv  3\langle H( t ){\rangle }^{2}$. It is then of interest to consider
the quantity $\langle \rho ( t ){\rangle }^{}/\rho ( t {)}_{{\rm crit}}$, and
we display this along with $\langle H(t)\rangle /H ( t {)}_{{\rm FRW}}$ in Fig.~(\ref{f5}).
We therefore effectively have $\rho /{\rho }_{{\rm crit}} <  1$ in a flat universe, and 
thus the oscillations are playing the role of dark energy. But an accelerated expansion
is not a result.

We now consider matter domination $w=0$. Here we find that the oscillations persist, but their
relative importance does not grow. For example the size of the oscillations in the quantity
 ${H}(t)^{2}/{H}(t)^{2}_{\rm FRW}$ does not grow.
This is reflected by the leveling off of $\rho(t)/\rho(t)_{\rm FRW}$ at some value determined by the
size of the oscillations. Both the initial and final value of $\rho(t)/\rho(t)_{\rm FRW}$ is adjustable
through the initial conditions.
At the same time $\langle H(t)\rangle /H ( t {)}_{{\rm FRW}}$ tends to unity.
The physically interesting quantity $\langle \rho ( t ){\rangle }^{}/\rho ( t {)}_{{\rm crit}}$
displays the least time dependence, and its essentially constant value, less than unity, is determined
solely by the size of the oscillations. Thus oscillations that have grown during a radiation dominated
era become ``frozen in'', in relative importance, during the matter dominated era.

Pressureless matter $w=0$ turns out to be the dividing line between the growth of the oscillations,
in relative terms, for $w>0$ and the
fading away of the oscillations for $w<0$. In particular a positive cosmological constant eventually
dominates the cosmological evolution and drives the effective $w$ negative. When this 
happens the oscillations fade away.

How natural it is to have the oscillations depends on how close to the Planck time the initial
conditions are specified. The initial conditions may be characterized, for example, by how much the acceleration
parameter deviates from the FRW value at that initial time. A given fractional deviation specified
at one initial time produces much less oscillation than the same fractional deviation specified
at an earlier initial time. Indeed at an initial time close to the Planck time $t_{\rm Pl}$, significant fine tuning is required
to produce a solution close to the FRW one. 

More appealing is to consider an initial time of order $30t_{\rm Pl}$.  This was used to
produce the $w=1/3$ results in Figs.~(\ref{f2},\ref{f4},\ref{f5}),
where a large fractional deviation, a factor of 10, in the acceleration parameter was used.
 For a small fractional deviation, say at the 10\% level,
the resulting oscillations are tiny, and they may not grow to be significant until cosmologically
interesting times have elapsed. This is based on numerical analysis that indicates that the quantity
$\langle \rho ( t ){\rangle }^{}/\rho ( t {)}_{{\rm crit}}$ in the radiation dominated era can be parameterized as
$1/( 1+c {\rm l}{\rm n}( t {)}^{a} )$ for small $c$ depending on the initial conditions and $a\sim 5$.
This is for times
 accessible to numerical analysis, whereas the era of radiation domination ends at $t\sim 10^{28}t_{\rm Pl}$,
and so whether this approximation remains true throughout this era is uncertain.
Clearly the main issue is to use the time of nucleosynthesis $t\sim 10^{18}t_{\rm Pl}$ to put a constraint on the deviations
from standard FRW evolution.

We have studied the higher derivative equations in a similar way, using the results of
the Appendix. This maintains energy conservation to ${\cal O}(\varepsilon^3)$,
and we have verified the insensitivity to the $\chi_{2n}$ parameters. We will
not display these numerical results, because they resemble
the results from the four derivative equations. The new qualitative feature is the existence of more than
one frequency mode of oscillation, where the relative size of the coefficients $c_{2n}$ determine how
easy it is to excite the various modes simultaneously. The resulting oscillations can deviate quite
dramatically from being pure sinusoidal
or even periodic. But for the observable smeared out quantities, the
results closely resemble those above.

\begin{figure}
\begin{center}
\includegraphics{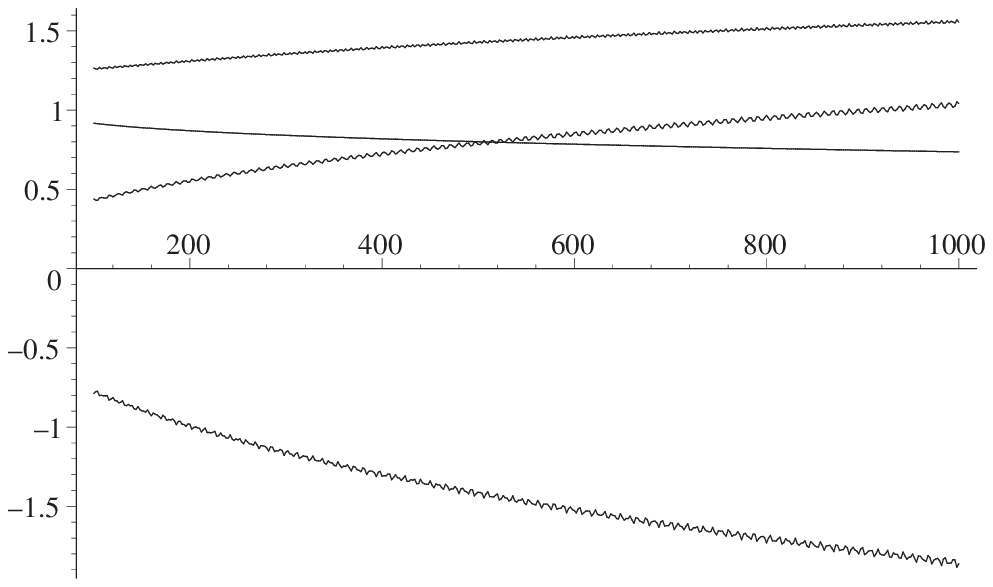}
\begin{picture}(0,0)(0,0)\small 
\put(0,0){4 derivatives}
\put(0,160){2 derivatives}
\put(0,135){6 derivatives}
\put(0,120){all}
\end{picture}
\end{center}
\caption{Contributions to $\rho(t)/\rho(t)_{\rm FRW}$ from terms with given numbers of
derivatives. The roughness of the curves is an artifact of the smearing.}
\label{f1} 
\end{figure}

We now look a little more closely at the origin of the result that $\rho(t)<\rho(t)_{\rm FRW}$.
In the pure oscillation case the modes with negative gravitational energy required negative
$\rho(t)$. Here it is clear that these modes are being excited and that they are responsible for driving
$\rho(t)<\rho(t)_{\rm FRW}$. In fact in the 4 derivative case there is only one oscillating mode
and it has negative gravitational energy.
At higher derivatives (\ref{e4}) suggests that contributions
to $\rho(t)$ alternate in sign depending on the order of derivatives. This is displayed in Fig.~(\ref{f1})
in the case of the equations with up to 6 derivatives and $c_6=1/5$.
The relative sizes of the contributions are consistent with (\ref{e4})
given that the lowest mode with $\omega=\omega_1$ dominates.
The conclusion is that we are linking a
dynamical instability in the expanding system to the growth of a negative 
component of the total gravitational energy.

As a side issue we mention that when the cosmological constant is negative there
appear to be solutions with oscillations on both large and small scales \cite{b}, where
the frequency of the large scale oscillation is proportional to $\sqrt{\Lambda}$.
Our results now allow us to determine the sign of $\rho$ needed to support these solutions
at any order in the number of
derivatives, and we find that it is typically negative.
 This is not surprising since the amplitude of the large time scale oscillations
can be adjusted to zero with appropriate initial conditions; this returns us to the pure oscillation
case we have treated above where we found negative $\rho(t) $ for some modes, in particular for the lowest frequency.
This leaves open the possibility that with appropriate initial conditions it may be
possible to excite the modes with positive energy in such a way as to have oscillations on large
as well as small time scales. The practical interest in this is unclear.

In this paper we have been concerned with a dynamical instability of standard
FRW expanding universes, where the matter energy density remains positive,
but is reduced in association
with the excitation of gravitation modes with negative energy. This is found to be a
generic feature of higher derivative gravity with any number of derivatives.
The phenomenological implications of this will have to be explored elsewhere,
but the instability appears to be of physical interest due to its slow logarithmic growth.
This instability is incompatible with a positive cosmological constant. This observation
may be interest in a theory in which the effective 4 dimensional cosmological constant
is a derived quantity, perhaps emerging from a higher dimensional theory \cite{a}.
The instability may provide a dynamical reason for the 4D flat spacetime to be preferred.
At the same time the oscillations and their gravitational coupling to matter may provide
a mechanism by which positive vacuum energy is converted through particle production
to matter energy.

\section*{Appendix}
We list here results for the quantities defined in (\ref{e8}-\ref{e9}), to the required orders
in the small amplitude expansion, and up to twelve derivatives.
\[
{{\cal{G}}_{p}}^{{(2)}}=({\textstyle\frac {
\partial ^{2}}{\partial t^{2}}}{A}(t)) + { 
\textstyle\frac {3}{4}} ({\textstyle\frac {\partial }{\partial t}}{A}(t))
^{2}
\]
\[
{{\cal{G}}_{\rho }}^{{(2)}}=
{ \textstyle\frac {3}{4}} ({\textstyle\frac {\partial }{\partial t}}
{A}(t))^{2}
\]
\[
{{\cal{G}}_{p}}^{{(4)}}=({\textstyle\frac {
\partial ^{4}}{\partial t^{4}}}{A}(t)) + 
{ \textstyle\frac {9}{4}} ({\textstyle\frac {\partial ^{2}}{\partial t
^{2}}}{A}(t))^{2} + 3({\textstyle\frac {\partial }{\partial t}}{A}(t))(
{\textstyle\frac {\partial ^{3}}{\partial t^{3}}}{A}(t)) + { 
\textstyle\frac {9}{4}} ({\textstyle\frac {\partial ^{2}}{\partial t^{2}}}
{A}(t))({\textstyle\frac {\partial }{\partial t}}{A}(t))
^{2}
\]
\[
{{\cal{G}}_{\rho }}^{{(4)}}=
 - { \textstyle\frac {3}{4}} ({\textstyle\frac {
\partial ^{2}}{\partial t^{2}}}{A}(t))^{2} + 
{ \textstyle\frac {3}{2}} ({\textstyle\frac {\partial }{\partial t}}
{A}(t))({\textstyle\frac {\partial ^{3}}{\partial t^{3}}}
{A}(t))+{ \textstyle\frac {9}{4}} ({\textstyle\frac {\partial ^{2}}{\partial t
^{2}}}{A}(t))({\textstyle\frac {\partial }{\partial t}}
{A}(t))^{2}
\]
\begin{eqnarray*}{{\cal{G}}_{p}}^{{(6)}}=({\textstyle\frac {
\partial ^{6}}{\partial t^{6}}}{A}(t)) + (
{ \textstyle\frac {27}{4}}  + {\chi _{6}})({\textstyle\frac {\partial 
^{3}}{\partial t^{3}}}{A}(t))^{2} + ({ 
\textstyle\frac {21}{2}}  + {\chi _{6}})({\textstyle\frac {\partial ^{2}}{\partial 
t^{2}}}{A}(t))({\textstyle\frac {\partial ^{4}}{\partial t^{4}}}
{A}(t)) \\
\mbox{} + { \textstyle\frac {9}{2}} ({\textstyle\frac {\partial }{
\partial t}}{A}(t))({\textstyle\frac {\partial ^{5}}{\partial t
^{5}}}{A}(t)) \end{eqnarray*}
\begin{eqnarray*}{{\cal{G}}_{\rho }}^{{(6)}}=
{ \textstyle\frac {3}{4}} ({\textstyle\frac {\partial ^{3}}{\partial t
^{3}}}{A}(t))^{2} - { \textstyle\frac {3}{2}} (
{\textstyle\frac {\partial ^{4}}{\partial t^{4}}}{A}(t))(
{\textstyle\frac {\partial ^{2}}{\partial t^{2}}}{A}(t)) + 
{ \textstyle\frac {3}{2}} ({\textstyle\frac {\partial }{\partial t}}
{A}(t))({\textstyle\frac {\partial ^{5}}{\partial t^{5}}}
{A}(t)) \\
\mbox{} + { \textstyle\frac {9}{2}} ({\textstyle\frac {\partial ^{4}}{
\partial t^{4}}}{A}(t))({\textstyle\frac {\partial }{\partial t
}}{A}(t))^{2} - ( - 9 - { \textstyle\frac {3}{2}} {
\chi _{6}})({\textstyle\frac {\partial }{\partial t}}{A}(t))(
{\textstyle\frac {\partial ^{3}}{\partial t^{3}}}{A}(t))(
{\textstyle\frac {\partial ^{2}}{\partial t^{2}}}{A}(t)) \\
\mbox{} - (3 + { \textstyle\frac {1}{2}} {\chi _{6}})(
{\textstyle\frac {\partial ^{2}}{\partial t^{2}}}{A}(t))^{3}\end{eqnarray*}
\begin{eqnarray*}{{\cal{G}}_{p}}^{{(8)}}=({\textstyle\frac {
\partial ^{8}}{\partial t^{8}}}{A}(t)) + (
{ \textstyle\frac {89}{4}}  + 6{\chi _{8}})({\textstyle\frac {
\partial ^{4}}{\partial t^{4}}}{A}(t))^{2} + (36 + 9{
\chi _{8}})({\textstyle\frac {\partial ^{3}}{\partial t^{3}}}{A}
(t))({\textstyle\frac {\partial ^{5}}{\partial t^{5}}}{A}(t))
 \\
\mbox{} + (19 + 3{\chi _{8}})({\textstyle\frac {\partial ^{2}}{
\partial t^{2}}}{A}(t))({\textstyle\frac {\partial ^{6}}{
\partial t^{6}}}{A}(t)) + 6({\textstyle\frac {\partial ^{7}}{
\partial t^{7}}}{A}(t))({\textstyle\frac {\partial }{\partial t
}}{A}(t))\end{eqnarray*}
\begin{eqnarray*}{{\cal{G}}_{\rho }}^{{(8)}}= - 
{ \textstyle\frac {3}{4}} ({\textstyle\frac {\partial ^{4}}{\partial t
^{4}}}{A}(t))^{2} + { \textstyle\frac {3}{2}} (
{\textstyle\frac {\partial ^{5}}{\partial t^{5}}}{A}(t))(
{\textstyle\frac {\partial ^{3}}{\partial t^{3}}}{A}(t)) - 
{ \textstyle\frac {3}{2}} ({\textstyle\frac {\partial ^{6}}{\partial t
^{6}}}{A}(t))({\textstyle\frac {\partial ^{2}}{\partial t^{2}}}
{A}(t)) \\
\mbox{} + { \textstyle\frac {3}{2}} ({\textstyle\frac {\partial ^{7}}{
\partial t^{7}}}{A}(t))({\textstyle\frac {\partial }{\partial t
}}{A}(t)) + { \textstyle\frac {27}{4}} ({\textstyle\frac {
\partial }{\partial t}}{A}(t))^{2}({\textstyle\frac {\partial ^{
6}}{\partial t^{6}}}{A}(t)) - ({ \textstyle\frac {69
}{4}}  + { \textstyle\frac {9}{2}} {\chi _{8}})({\textstyle\frac {
\partial ^{2}}{\partial t^{2}}}{A}(t))^{2}({\textstyle\frac {
\partial ^{4}}{\partial t^{4}}}{A}(t)) \\
\mbox{} - ( - { \textstyle\frac {69}{4}}  - { 
\textstyle\frac {9}{2}} {\chi _{8}})({\textstyle\frac {\partial }{\partial t}}
{A}(t))({\textstyle\frac {\partial ^{2}}{\partial t^{2}}}
{A}(t))({\textstyle\frac {\partial ^{5}}{\partial t^{5}}}
{A}(t)) \\
\mbox{} - ( - { \textstyle\frac {69}{2}}  - 9{\chi _{8}})(
{\textstyle\frac {\partial }{\partial t}}{A}(t))({\textstyle\frac {
\partial ^{3}}{\partial t^{3}}}{A}(t))({\textstyle\frac {
\partial ^{4}}{\partial t^{4}}}{A}(t)) \end{eqnarray*}
\begin{eqnarray*}{{\cal{G}}_{p}}^{{(10)}}=({\textstyle\frac {
\partial ^{10}}{\partial t^{10}}}{A}(t)) + (
{ \textstyle\frac {223}{4}}  + {\chi _{10}})({\textstyle\frac {
\partial ^{5}}{\partial t^{5}}}{A}(t))^{2} + (
{ \textstyle\frac {191}{2}}  + { \textstyle\frac {8}{5}} 
{\chi _{10}})({\textstyle\frac {\partial ^{4}}{\partial t^{4}}}{
A}(t))({\textstyle\frac {\partial ^{6}}{\partial t^{6}}}{A}(t))
 \\
\mbox{} + (59 + { \textstyle\frac {4}{5}} {\chi _{10}})(
{\textstyle\frac {\partial ^{3}}{\partial t^{3}}}{A}(t))(
{\textstyle\frac {\partial ^{7}}{\partial t^{7}}}{A}(t)) + (26 + 
{ \textstyle\frac {1}{5}} {\chi _{10}})({\textstyle\frac {\partial 
^{2}}{\partial t^{2}}}{A}(t))({\textstyle\frac {\partial ^{8}}{
\partial t^{8}}}{A}(t)) \\
\mbox{} + { \textstyle\frac {15}{2}} ({\textstyle\frac {\partial }{
\partial t}}{A}(t))({\textstyle\frac {\partial ^{9}}{\partial t
^{9}}}{A}(t)) \end{eqnarray*}
\begin{eqnarray*}{{\cal{G}}_{\rho }}^{{(10)}}=
{ \textstyle\frac {3}{4}} ({\textstyle\frac {\partial ^{5}}{\partial t
^{5}}}{A}(t))^{2} - { \textstyle\frac {3}{2}} (
{\textstyle\frac {\partial ^{4}}{\partial t^{4}}}{A}(t))(
{\textstyle\frac {\partial ^{6}}{\partial t^{6}}}{A}(t)) + 
{ \textstyle\frac {3}{2}} ({\textstyle\frac {\partial ^{3}}{\partial t
^{3}}}{A}(t))({\textstyle\frac {\partial ^{7}}{\partial t^{7}}}
{A}(t)) \\
\mbox{} - { \textstyle\frac {3}{2}} ({\textstyle\frac {\partial ^{8}}{
\partial t^{8}}}{A}(t))({\textstyle\frac {\partial ^{2}}{
\partial t^{2}}}{A}(t)) + { \textstyle\frac {3}{2}} 
({\textstyle\frac {\partial }{\partial t}}{A}(t))({\textstyle\frac {
\partial ^{9}}{\partial t^{9}}}{A}(t)) \\
\mbox{} - ( - { \textstyle\frac {165}{2}}  - { 
\textstyle\frac {3}{2}} {\chi _{10}})({\textstyle\frac {\partial ^{4}}{\partial t
^{4}}}{A}(t))({\textstyle\frac {\partial }{\partial t}}
{A}(t))({\textstyle\frac {\partial ^{5}}{\partial t^{5}}}
{A}(t)) \\
\mbox{} - ( - 63 - { \textstyle\frac {9}{10}} {\chi _{10}})
({\textstyle\frac {\partial ^{3}}{\partial t^{3}}}{A}(t))(
{\textstyle\frac {\partial ^{6}}{\partial t^{6}}}{A}(t))(
{\textstyle\frac {\partial }{\partial t}}{A}(t)) \\
\mbox{} - ( - { \textstyle\frac {33}{2}}  - { 
\textstyle\frac {3}{10}} {\chi _{10}})({\textstyle\frac {\partial ^{4}}{\partial 
t^{4}}}{A}(t))({\textstyle\frac {\partial ^{3}}{\partial t^{3}}}
{A}(t))^{2} \\
\mbox{} - ( - { \textstyle\frac {93}{4}}  - { 
\textstyle\frac {3}{10}} {\chi _{10}})({\textstyle\frac {\partial }{\partial t}}
{A}(t))({\textstyle\frac {\partial ^{7}}{\partial t^{7}}}
{A}(t))({\textstyle\frac {\partial ^{2}}{\partial t^{2}}}
{A}(t)) + 9({\textstyle\frac {\partial ^{8}}{\partial t^{8}}}
{A}(t))({\textstyle\frac {\partial }{\partial t}}{A}(t))
^{2} \\
\mbox{} - (33 + { \textstyle\frac {3}{5}} {\chi _{10}})(
{\textstyle\frac {\partial ^{4}}{\partial t^{4}}}{A}(t))^{2}(
{\textstyle\frac {\partial ^{2}}{\partial t^{2}}}{A}(t)) - (
{ \textstyle\frac {33}{2}}  + { \textstyle\frac {3}{10}} 
{\chi _{10}})({\textstyle\frac {\partial ^{3}}{\partial t^{3}}}{
A}(t))({\textstyle\frac {\partial ^{5}}{\partial t^{5}}}{A}(t))
({\textstyle\frac {\partial ^{2}}{\partial t^{2}}}{A}(t)) \\
\mbox{} - ({ \textstyle\frac {93}{4}}  + { 
\textstyle\frac {3}{10}} {\chi _{10}})({\textstyle\frac {\partial ^{2}}{\partial 
t^{2}}}{A}(t))^{2}({\textstyle\frac {\partial ^{6}}{\partial t^{
6}}}{A}(t)) \end{eqnarray*}

\begin{eqnarray*}{{\cal{G}}_{p}}^{{(12)}}=({\textstyle\frac {
\partial ^{12}}{\partial t^{12}}}{A}(t)) + (318 + {\chi 
_{12}})({\textstyle\frac {\partial ^{6}}{\partial t^{6}}}{A}(t))
^{2} + (549 + { \textstyle\frac {8}{5}} {\chi _{12}})(
{\textstyle\frac {\partial ^{5}}{\partial t^{5}}}{A}(t))(
{\textstyle\frac {\partial ^{7}}{\partial t^{7}}}{A}(t)) \\
\mbox{} + ({ \textstyle\frac {1395}{4}}  + { 
\textstyle\frac {29}{35}} {\chi _{12}})({\textstyle\frac {\partial ^{4}}{
\partial t^{4}}}{A}(t))({\textstyle\frac {\partial ^{8}}{
\partial t^{8}}}{A}(t)) + ({ \textstyle\frac {315}{2}
}  + { \textstyle\frac {2}{7}} {\chi _{12}})({\textstyle\frac {
\partial ^{3}}{\partial t^{3}}}{A}(t))({\textstyle\frac {
\partial ^{9}}{\partial t^{9}}}{A}(t)) \\
\mbox{} + (48 + { \textstyle\frac {2}{35}} {\chi _{12}})(
{\textstyle\frac {\partial ^{2}}{\partial t^{2}}}{A}(t))(
{\textstyle\frac {\partial ^{10}}{\partial t^{10}}}{A}(t)) + 9(
{\textstyle\frac {\partial }{\partial t}}{A}(t))({\textstyle\frac {
\partial ^{11}}{\partial t^{11}}}{A}(t))\end{eqnarray*}
\begin{eqnarray*}{{\cal{G}}_{\rho }}^{{(12)}}= - 
{ \textstyle\frac {3}{4}} ({\textstyle\frac {\partial ^{6}}{\partial t
^{6}}}{A}(t))^{2} + { \textstyle\frac {3}{2}} (
{\textstyle\frac {\partial ^{5}}{\partial t^{5}}}{A}(t))(
{\textstyle\frac {\partial ^{7}}{\partial t^{7}}}{A}(t)) - 
{ \textstyle\frac {3}{2}} ({\textstyle\frac {\partial ^{4}}{\partial t
^{4}}}{A}(t))({\textstyle\frac {\partial ^{8}}{\partial t^{8}}}
{A}(t)) \\
\mbox{} + { \textstyle\frac {3}{2}} ({\textstyle\frac {\partial ^{3}}{
\partial t^{3}}}{A}(t))({\textstyle\frac {\partial ^{9}}{
\partial t^{9}}}{A}(t)) - { \textstyle\frac {3}{2}} 
({\textstyle\frac {\partial ^{10}}{\partial t^{10}}}{A}(t))(
{\textstyle\frac {\partial ^{2}}{\partial t^{2}}}{A}(t)) + 
{ \textstyle\frac {3}{2}} ({\textstyle\frac {\partial }{\partial t}}
{A}(t))({\textstyle\frac {\partial ^{11}}{\partial t^{11}}}
{A}(t)) \\
\mbox{} - ( - { \textstyle\frac {3825}{8}}  - { 
\textstyle\frac {3}{2}} {\chi _{12}})({\textstyle\frac {\partial }{\partial t}}
{A}(t))({\textstyle\frac {\partial ^{5}}{\partial t^{5}}}
{A}(t))({\textstyle\frac {\partial ^{6}}{\partial t^{6}}}
{A}(t)) \\
\mbox{} - ( - { \textstyle\frac {207}{4}}  - { 
\textstyle\frac {3}{35}} {\chi _{12}})({\textstyle\frac {\partial }{\partial t}}
{A}(t))({\textstyle\frac {\partial ^{9}}{\partial t^{9}}}
{A}(t))({\textstyle\frac {\partial ^{2}}{\partial t^{2}}}
{A}(t)) \\
\mbox{} - ({ \textstyle\frac {315}{4}}  + { 
\textstyle\frac {6}{35}} {\chi _{12}})({\textstyle\frac {\partial ^{3}}{\partial 
t^{3}}}{A}(t))({\textstyle\frac {\partial ^{7}}{\partial t^{7}}}
{A}(t))({\textstyle\frac {\partial ^{2}}{\partial t^{2}}}
{A}(t)) \\
\mbox{} - ({ \textstyle\frac {2115}{8}}  + { 
\textstyle\frac {51}{70}} {\chi _{12}})({\textstyle\frac {\partial ^{4}}{
\partial t^{4}}}{A}(t))({\textstyle\frac {\partial ^{6}}{
\partial t^{6}}}{A}(t))({\textstyle\frac {\partial ^{2}}{
\partial t^{2}}}{A}(t)) + { \textstyle\frac {45}{4}} 
({\textstyle\frac {\partial ^{10}}{\partial t^{10}}}{A}(t))(
{\textstyle\frac {\partial }{\partial t}}{A}(t))^{2} \\
\mbox{} - ({ \textstyle\frac {207}{4}}  + { 
\textstyle\frac {3}{35}} {\chi _{12}})({\textstyle\frac {\partial ^{2}}{\partial 
t^{2}}}{A}(t))^{2}({\textstyle\frac {\partial ^{8}}{\partial t^{
8}}}{A}(t)) - ({ \textstyle\frac {855}{8}}  + 
{ \textstyle\frac {27}{70}} {\chi _{12}})({\textstyle\frac {
\partial ^{5}}{\partial t^{5}}}{A}(t))^{2}({\textstyle\frac {
\partial ^{2}}{\partial t^{2}}}{A}(t)) \\
\mbox{} - ( - { \textstyle\frac {855}{8}}  - { 
\textstyle\frac {27}{70}} {\chi _{12}})({\textstyle\frac {\partial ^{3}}{
\partial t^{3}}}{A}(t))({\textstyle\frac {\partial ^{5}}{
\partial t^{5}}}{A}(t))({\textstyle\frac {\partial ^{4}}{
\partial t^{4}}}{A}(t)) \\
\mbox{} - ( - { \textstyle\frac {315}{4}}  - { 
\textstyle\frac {6}{35}} {\chi _{12}})({\textstyle\frac {\partial ^{3}}{\partial 
t^{3}}}{A}(t))^{2}({\textstyle\frac {\partial ^{6}}{\partial t^{
6}}}{A}(t)) \\
\mbox{} - ( - { \textstyle\frac {2745}{8}}  - { 
\textstyle\frac {9}{10}} {\chi _{12}})({\textstyle\frac {\partial }{\partial t}}
{A}(t))({\textstyle\frac {\partial ^{7}}{\partial t^{7}}}
{A}(t))({\textstyle\frac {\partial ^{4}}{\partial t^{4}}}
{A}(t)) \\
\mbox{} - ( - { \textstyle\frac {729}{4}}  - { 
\textstyle\frac {12}{35}} {\chi _{12}})({\textstyle\frac {\partial }{\partial t}}
{A}(t))({\textstyle\frac {\partial ^{8}}{\partial t^{8}}}
{A}(t))({\textstyle\frac {\partial ^{3}}{\partial t^{3}}}
{A}(t)) - ({ \textstyle\frac {285}{8}}  + 
{ \textstyle\frac {9}{70}} {\chi _{12}})({\textstyle\frac {\partial 
^{4}}{\partial t^{4}}}{A}(t))^{3}\end{eqnarray*}

\section*{Acknowledgments}
This research was supported in part by the Natural Sciences 
and Engineering Research Council of Canada.

\end{document}